\documentclass[sigconf,nonacm]{acmart}

\usepackage{booktabs}
\usepackage{multirow}
\usepackage{subcaption}
\usepackage{algorithm}
\usepackage{algpseudocode}
\usepackage{xcolor}
\usepackage{colortbl}
\usepackage{amsmath}
\usepackage{placeins}
\usepackage{balance}
\usepackage{needspace}

\newcommand{\K}{\mathcal{K}}
\newcommand{\D}{\mathcal{D}}
\newcommand{\Ch}{\hat{C}}

\setcopyright{none}
\begin{document}

\title[CASH: A Capability-Aware Subsidy Harness]{CASH: A Capability-Aware Subsidy Harness for Advertiser Subsidy Allocation}

\author{Minmao Wang}
\authornote{This work was completed during an internship at Kuaishou Technology.}
\affiliation{%
  \institution{Fudan University}
  \city{Shanghai}
  \country{China}
}
\email{mmwang25@m.fudan.edu.cn}

\author{Huihuang Lin}
\affiliation{%
  \institution{Kuaishou Technology}
  \city{Beijing}
  \country{China}
}
\email{linhuihuang@kuaishou.com} 

\author{Likang Wu}
\authornote{Corresponding authors.}
\affiliation{%
  \institution{Tianjin University}
  \city{Tianjin}
  \country{China}
}
\email{wulk@tju.edu.cn}

\author{Hongke Zhao}
\affiliation{%
  \institution{Tianjin University}
  \city{Tianjin}
  \country{China}
}
\email{hongke@tju.edu.cn}

\author{Shijie Yi}
\authornotemark[2]
\affiliation{%
  \institution{Kuaishou Technology}
  \city{Beijing}
  \country{China}
}
\email{yishijie@kuaishou.com}

\author{Fei Pan}
\affiliation{%
  \institution{Kuaishou Technology}
  \city{Beijing}
  \country{China}
}
\email{panfei05@kuaishou.com}

\renewcommand{\shortauthors}{Wang et al.}

\begin{abstract}
Advertising platforms subsidize advertisers through billing discounts, and \emph{Advertiser Subsidy Allocation} (ASA) chooses a subsidy level for every account to maximize total discounted revenue under a platform-wide subsidy-rate constraint. In practice, operators configure these subsidies from business rules, advertiser tiers and operating goals. Such configuration draws on rich business experience, yet as the advertiser base grows, tier-level policies can neither resolve how individual accounts respond across subsidy levels nor allocate a shared budget at the granularity of single accounts. We propose CASH, a Capability-Aware Subsidy Harness for ASA. CASH pairs a knowledge-guided large language model (LLM) with a data-driven numerical response model, learns from realized outcomes where each is locally reliable across accounts and subsidy levels, and turns this reliability into explicit capability licenses that govern source selection, agent review and budget allocation. Its \emph{Expected Policy Coverage Gain} (EPCG) directs a limited feedback budget toward evidence expected to expand high-value reliable coverage, and the resulting feedback further improves the numerical model and updates capability estimates. In a one-month online experiment with CASH serving 60\% of a large-scale advertising platform's traffic, CASH raises revenue per account by 15.35\% and ADVV by 19.86\% over Control at an essentially unchanged subsidy rate, with further gains in revenue quality, LTV and retention.
\end{abstract}


\maketitle

\section{Introduction}

\begin{figure}[t]
\centering
\includegraphics[width=\columnwidth]{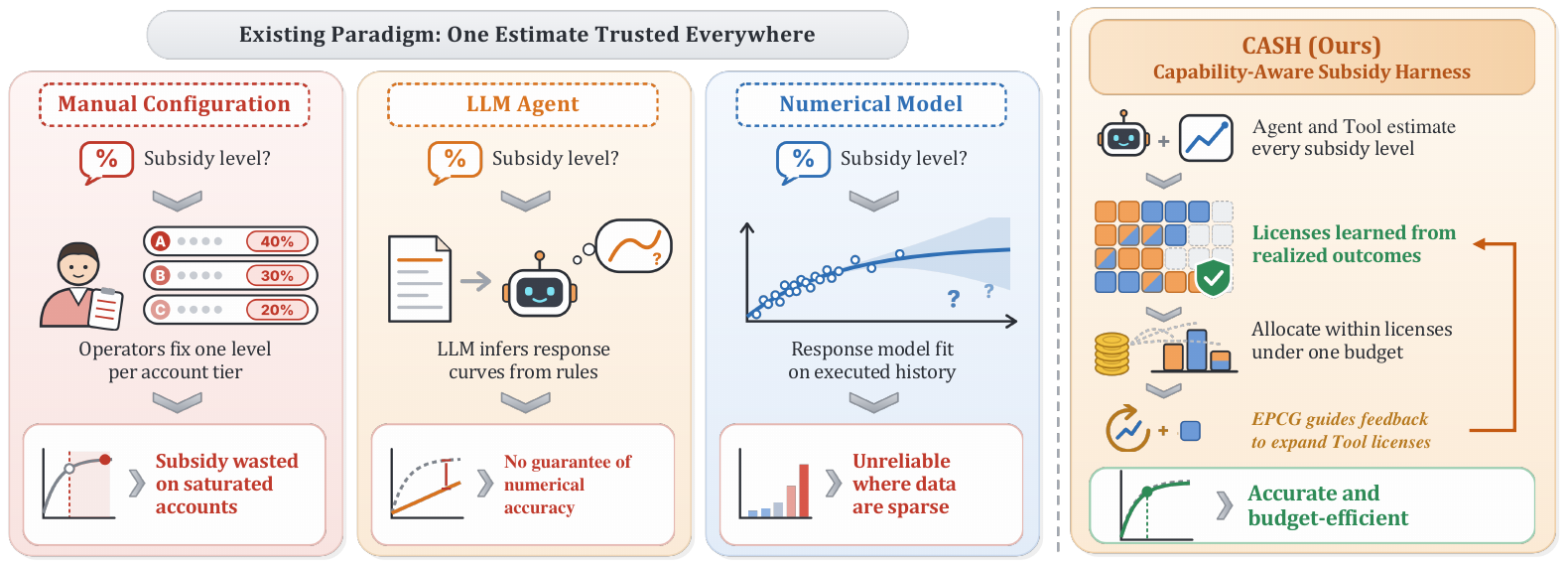}
\caption{From manual configuration and single-source estimation to capability-aware subsidy allocation. CASH combines agent and numerical estimates through outcome-grounded licenses under a shared budget.}
\Description{Comparison of manual configuration, prediction-agent estimation and numerical modeling with CASH. CASH uses outcome-grounded capability licenses to constrain shared-budget allocation and updates the licenses through feedback.}
\label{fig:teaser}
\end{figure}

Online advertising platforms~\cite{yang2026platformbid,han2026autobidding} connect vast numbers of users and advertisers, and their long-term growth depends on effective traffic matching, a thriving advertiser ecosystem and sustainable commercial revenue. Yet an advertiser's competitiveness in impression-level auctions does not always reflect its business potential. Advertisers with high conversion potential may receive insufficient exposure due to cold start, sparse historical feedback or weak ranking competitiveness, hindering their growth and long-term retention~\cite{ye2023coldstart}. Platforms therefore provide differentiated account-level support by raising ranking bids through an uplift factor while discounting billing, helping promising advertisers become more competitive and obtain additional traffic. Advertisers, however, respond very differently to the same support. Excessive subsidies increase platform cost without commensurate revenue gains, whereas insufficient subsidies constrain the growth of promising advertisers. Efficiently allocating a limited subsidy budget is therefore a central challenge in advertising operations.

In industrial systems, account-level subsidies are configured mainly through manual rules and operator experience, such as adjusting discounts during promotions or helping new advertisers ramp up. These practices cannot capture each account's marginal revenue response at different subsidy levels. For example, when a 30\% discount already secures sufficient traffic, deepening it to 50\% may add little revenue while substantially increasing subsidy cost. Subsidies also affect competition for traffic, motivating a platform-level allocation objective. We formulate this problem as \emph{Advertiser Subsidy Allocation} (ASA), which selects a subsidy level for each account to maximize total platform revenue under a platform-wide subsidy-rate constraint. ASA operates at the account level and shapes subsequent traffic allocation through the existing auction mechanism.

Large language models (LLMs)~\cite{zhao2026llmsurvey,brown2020fewshot} offer a natural way to scale subsidy decisions by incorporating business rules, account operating information and operational knowledge into revenue-response estimation. Their estimates, however, do not guarantee numerical accuracy for specific subsidy actions~\cite{tan2024language,song2026dara}. Data-driven numerical response models provide an alternative, learning directly from realized outcomes and predicting efficiently where historical feedback is abundant. Yet subsidies concentrate on a small number of accounts and levels, leaving many account-action combinations poorly supported. Neither estimator is uniformly reliable; their strengths depend on account states, subsidy levels and available evidence, as shown in Figure~\ref{fig:teaser}.

The central challenge is to make heterogeneous estimators collaborate reliably while allowing their usable decision range to evolve with experience. In ASA, an error at a single account-action pair can alter the allocation of a shared budget and consequently change decisions for other accounts. Meanwhile, the evidence available to assess reliability is shaped by previous allocation decisions, so poorly supported actions may remain difficult to evaluate. Reliable allocation thus requires both outcome-grounded limits on where each estimator can be used and a principled way to acquire evidence that expands these limits over time.

We propose \textbf{CASH} (Capability-Aware Subsidy Harness), which integrates estimator collaboration, reliability control and shared-budget allocation through explicit capability boundaries, as shown in Figure~\ref{fig:framework}. CASH pairs an LLM-based prediction agent with a numerical response model to estimate account-level revenue curves. A capability plugin evaluates both sources using realized outcomes and issues per-level licenses for individual account states. These licenses constrain routine source selection and the candidate actions available to the shared-budget allocator. A separate LLM-based orchestrating agent reviews contested accounts using the predictions, capability report and operating information, choosing among licensed sources or withholding the subsidy. As new feedback arrives, CASH updates these licenses and adjusts the division of labor between the prediction agent and the numerical model.

CASH further introduces \emph{Expected Policy Coverage Gain} (EPCG) to manage how these capability boundaries evolve. Because only executed actions yield feedback, historical logs leave gaps in the evidence needed to assess alternative decisions. EPCG allocates a limited feedback quota according to each region's expected contribution to future value-weighted reliable coverage of the numerical model, measured through changes in regional admission while holding other license conditions fixed. Acquired feedback supports model retraining and capability reassessment, and CASH updates or revokes licenses as evidence changes, while EPCG prioritizes the interventions expected to expand reliable Tool coverage. EPCG thereby connects evidence acquisition to the same capability boundaries that govern current allocation, closing the loop between decision making and capability evolution.

Our main contributions are as follows.
\begin{itemize}
\item \textbf{Problem formulation.} To our knowledge, we are the first to study advertiser subsidy allocation from an advertising operations perspective, formulating it as account-level optimization under a platform-wide subsidy-rate constraint. We identify its core challenges: heterogeneous estimator reliability across accounts and subsidy levels, and cross-account error propagation through the shared budget.

\item \textbf{Capability-aware harness.} We propose CASH, which learns outcome-grounded, per-level capability licenses for a prediction agent and a numerical response model, and uses these licenses to govern source selection and shared-budget allocation. EPCG directs limited feedback toward regions with the greatest expected expansion of reliable coverage, closing the loop between evidence acquisition and capability evolution.

\item \textbf{Industrial validation.} In a one-month online experiment with CASH serving 60\% of platform traffic, CASH increases revenue per account by 15.35\% and ADVV by 19.86\% over Control at an essentially unchanged subsidy rate, while improving revenue quality, LTV, and retention.
\end{itemize}

\section{Problem Formulation}
\label{sec:problem}

In each decision cycle, the platform allocates subsidies to a set of advertiser accounts $\mathcal I$. The state $s_i$ of account $i$ consists of its historical behavior, historical subsidy actions and available operating information visible before the decision, and the platform selects a discount factor $k_i$ from the subsidy grid $\K$, where a smaller $k$ means a larger discount, or takes the separate no-discount action $k_i=1$. Let $Q(s,k)$ denote the expected pre-discount revenue under state $s$ and action $k$. Expected discounted revenue and subsidy are
\begin{equation}
C(s,k)=kQ(s,k),\qquad S(s,k)=\tfrac{1-k}{k}\,C(s,k),
\end{equation}
and the realized discounted revenue is denoted by $y$. ASA maximizes expected total discounted revenue under a platform-wide subsidy-rate constraint,
\begin{equation}
\label{eq:asa}
\begin{aligned}
\max_{\{k_i\}}\quad &\sum_{i\in\mathcal I}C(s_i,k_i)\\
\text{s.t.}\quad &\rho_{\min}\le
\frac{\sum_i S(s_i,k_i)}{\sum_i C(s_i,k_i)}\le\rho_{\max},\\
&k_i\in\K_i.
\end{aligned}
\end{equation}
where the subsidy rate is the ratio of total subsidy to total discounted revenue and the executable levels $\K_i\subseteq\K\cup\{1\}$ are determined in \S\ref{sec:route}, with $\K_i=\{1\}$ for \textsc{Fallback}. Following prior incentive-allocation work~\cite{ai2022lbcf,sun2024e3ir}, we model responses at the account level under recent market conditions and treat $\mathbb E[y\mid s,k]$ learned from logs as the interventional response, assuming that $s$ contains the information behind historical subsidy decisions so that, given $s$, the historical level is independent of potential outcomes. The support condition of \S\ref{sec:cap} further restricts each estimator to state and level pairs with sufficient historical feedback. $C$ is supplied by two estimators, a numerical response model $\Ch^{T}$ and a prediction agent $\Ch^{A}$. Historical subsidies concentrate on a few accounts and levels, so the reliability of both estimators varies markedly across~$(s,k)$.

With non-negative multipliers $\mu=(\mu_{\min},\mu_{\max})$ for the two sides of the constraint, the Lagrangian subproblem of Eq.~\eqref{eq:asa} decomposes over accounts into $\max_{k\in\K_i}a_k(\mu)C(s_i,k)$ with shared coefficient $a_k(\mu)=1+(\mu_{\min}-\mu_{\max})\frac{1-k}{k}-\mu_{\min}\rho_{\min}+\mu_{\max}\rho_{\max}$.

\begin{proposition}
\label{prop:scale}
For given dual prices and executable set $\K_i$, if $C'(s_i,\cdot)=\zeta\,C(s_i,\cdot)$ with $\zeta>0$, then
\begin{equation}
\operatorname*{arg\,max}_{k\in\K_i}a_k(\mu)C'(s_i,k)=\operatorname*{arg\,max}_{k\in\K_i}a_k(\mu)C(s_i,k),
\end{equation}
whereas the account's constraint contribution, \mbox{$\big(\tfrac{1-k}{k}-\rho\big)C'(s_i,k)$} with $\rho\in\{\rho_{\min},\rho_{\max}\}$, scales with $\zeta$.
\end{proposition}

At fixed dual prices a scale bias thus leaves an account's choice unchanged, but changes its budget contribution, which can alter the optimal dual prices and other accounts' allocations, so where each estimator applies must be decided before allocation. The log $\D_t=\{(s_j,k_j,y_j)\}$ contains outcomes only for executed actions. Allocation therefore determines which new evidence becomes available to reassess each estimator's reliable range.

\section{CASH: A Capability-Aware Subsidy Harness}
\label{sec:method}

\begin{figure*}[t]
\centering
\includegraphics[width=\textwidth]{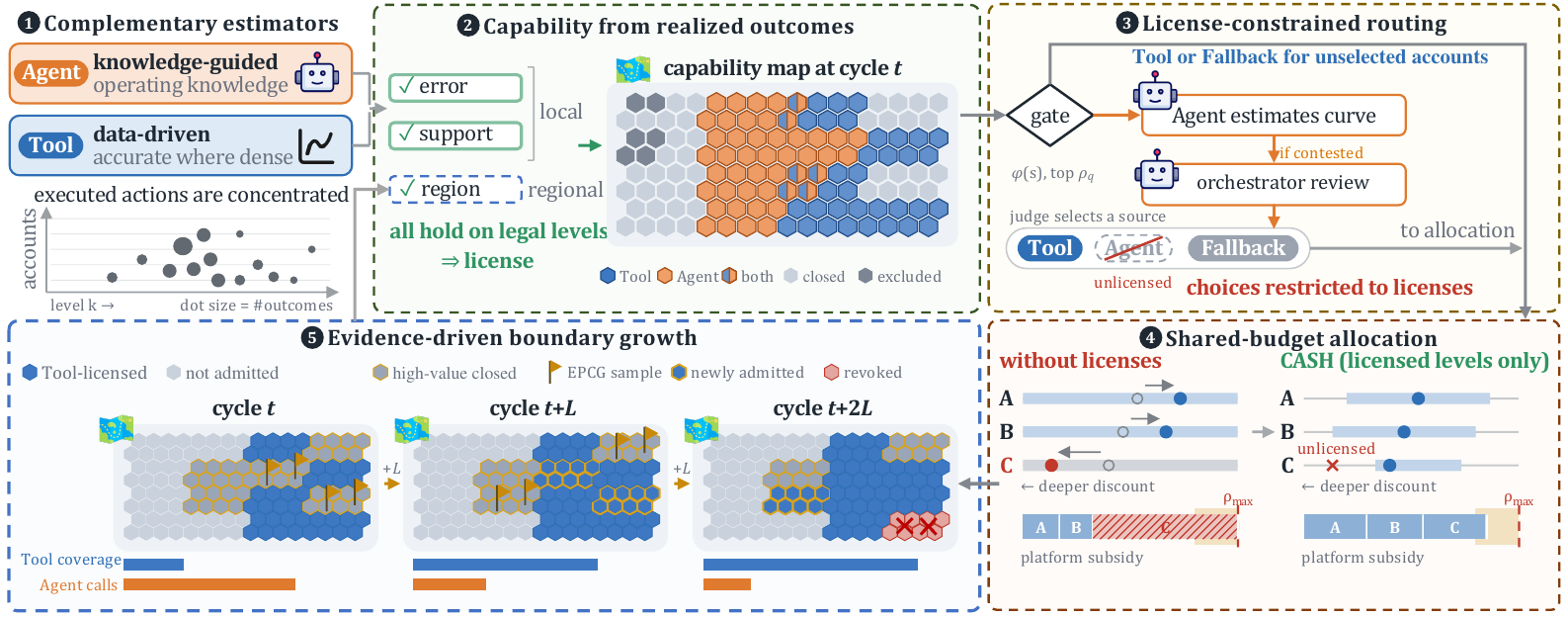}
\caption{Overview of CASH. The numerical model and prediction agent provide complementary revenue estimates. Outcome-grounded licenses combine local error, empirical support and regional admission to guide source selection, review by the orchestrating agent and routine allocation under a shared subsidy-rate constraint. EPCG prioritizes feedback by expected gains in value-weighted Tool coverage, while matured outcomes update admission and support model retraining.}
\Description{CASH combines Agent and Tool estimates, checks local and regional reliability to grant licenses, constrains source selection and shared-budget allocation, and uses EPCG-guided feedback to expand or revoke Tool coverage.}
\label{fig:framework}
\end{figure*}

CASH links response estimation, capability assessment and resource allocation in a feedback loop, as shown in Figure~\ref{fig:framework}. The capability plugin uses matured outcomes to determine each source's licensed levels. Routing selects a source within these licenses, and the allocator assigns levels under the shared budget. EPCG selects evidence acquisition opportunities by their expected contribution to future Tool coverage; exploration configurations pass separate business checks and share the same budget. Matured feedback updates regional admission and supports scheduled model retraining, determining the licenses available in subsequent cycles. Algorithm~\ref{alg:loop} summarizes one decision cycle.

\subsection{Complementary Response Estimation}
\label{sec:est}

CASH combines a numerical response model (Tool) learned from historical interventions with a prediction agent (Agent) that uses account history, operating information and business rules. Both predict discounted revenue on the same scale.

Tool predicts pre-discount revenue $f_\theta(s,k)$, giving $\Ch^{T}(s,k)=k\,f_\theta(s,k)$. It trains on matured feedback $\D_t$ with rolling loss $\ell\big(k_jf_\theta(s_j,k_j),y_j\big)$. Each logged prediction uses the model version preceding the inclusion of its feedback in training, keeping downstream errors out-of-sample. Agent reads state $s$, operating information $\xi_s$ and subsidy rules $\mathcal R$, and returns all levels in one call: $\Ch^{A}(s,\cdot)=\mathrm{LLM}\big(\mathrm{prompt}(s,\xi_s,\mathcal R)\big)$. Calls follow the routine quota in \S\ref{sec:route}, with additional random audits.

Optimizing noisy curves favors over-estimated levels. Since deeper discounts raise ranking bids through the uplift factor, pre-discount revenue should not decrease. We therefore project $\hat Q^{T}=f_\theta$ and $\hat Q^{A}=\Ch^{A}/k$ onto non-increasing sequences in $k$ by isotonic regression, then multiply by $k$. Capability estimation and allocation use these projected curves.

\subsection{Outcome-Grounded Capability Licenses}
\label{sec:cap}

The capability plugin assesses source reliability from realized prediction errors. Errors become observable when feedback matures for valid predictions made with decision-time information.

For each source $m\in\{T,A\}$, we fit an error-scale model $\hat g^m$ to predict relative error for different account states and subsidy levels. Its target is $e^m_j=|y_j-\Ch^m_j|/\max(\Ch^m_j,c_0)$, where $\Ch^m_j=\Ch^m(s_j,k_j)$ and $c_0>0$ stabilizes the revenue scale. A disjoint, chronologically later calibration set $\D^{\mathrm{cal},m}$ provides a quantile of normalized errors~\cite{lei2018distribution}. This quantile rescales the predicted relative error into a relative bound, which we convert to a monetary bound using the predicted revenue scale:
\begin{equation}
\label{eq:bound}
\begin{aligned}
\tilde e^m_j&=\frac{e^m_j}{\max\big(\hat g^m(s_j,k_j),\epsilon_g\big)},\\
\lambda^m&=\tilde e^m_{(r^m_{\mathrm{cal}})},\\[2pt]
\bar u^m(s,k)&=\lambda^m\max\big(\hat g^m(s,k),\epsilon_g\big),\\
u^m(s,k)&=\bar u^m(s,k)\max\big(\Ch^m(s,k),c_0\big),
\end{aligned}
\end{equation}
Here $n^m_{\mathrm{cal}}=|\D^{\mathrm{cal},m}|$, $r^m_{\mathrm{cal}}=\lceil(n^m_{\mathrm{cal}}+1)(1-\alpha)\rceil$, and $\tilde e^m_{(i)}$ is the $i$-th ordered score, taken as $\infty$ for $i>n^m_{\mathrm{cal}}$. The floor $\epsilon_g>0$ stabilizes normalization; nominal coverage is $1-\alpha$. Agent calibration uses valid routine-call predictions ($M(s)=1$, \S\ref{sec:route}); audits feed regional admission and the judge.

The error-scale model is trained on executed $(s,k)$, and its use at other levels is restricted to neighborhoods with sufficient empirical support. We measure support by kernel-weighted matured feedback, capping each account's contribution so that correlated feedback from one account is not counted as independent evidence,
\begin{equation}
\label{eq:support}
\sigma(s,k)=\sum_{i'}\min\Big(1,\!\!\sum_{j\in\D^H_t:\,i(j)=i'}\!\!\kappa_s(s,s_j)\,\kappa_k(k,k_j)\Big),
\end{equation}
where $i(j)$ is the account of sample $j$, and $\kappa_s$ and $\kappa_k$ are bounded-support kernels over account states and levels (forms and bandwidths in Appendix~\ref{app:impl}). $\D^H_t\subseteq\D_t$ is the feedback matured within the last $H$ cycles, the same evidence window as \S\ref{sec:evolve}, so outdated evidence stops supporting a license.

For source $m$ at state $s$, source-specific thresholds and regional admission define the licensed levels:
\begin{equation}
\label{eq:license}
\begin{aligned}
\mathcal L^m(s)=\Big\{k\in\K^{\mathrm{safe}}(s):{}&\\[-3pt]
\bar u^m(s,k)\le\eta^m_u,\ \sigma(s,k)\ge\eta^m_\sigma,\ A^m_{r(s,k)}=1\Big\},
\end{aligned}
\end{equation}
where $\K^{\mathrm{safe}}(s)$ is the set of levels confirmed legal by business rules (\S\ref{sec:route}) and $A^m_r$ is the admission state of region $r$ (\S\ref{sec:evolve}). All three conditions must hold for a level to be licensed. The plugin returns the licensed levels together with error and support estimates in the capability report of \S\ref{sec:route}.

\subsection{Capability-Constrained Routing}
\label{sec:route}

Let $\K^{\mathrm{safe}}(s)\subseteq\K$ be the levels confirmed legal by business rules, with which every license is intersected. Not every account needs a discount. When no source can support a decision, the account takes \textsc{Fallback} and receives no discount in the current cycle ($k=1$), so it keeps its normal billing and adds no subsidy.

Using pre-call account information, a judge $h_\phi$ gives an account-level source-preference score $p_A(s)=h_\phi(s)$. It is trained on error comparisons over executed actions, with samples weighted by the error gap to emphasize costly source-selection mistakes,
\begin{equation}
\label{eq:judge}
\begin{aligned}
\phi&=\operatorname*{arg\,min}_\phi\sum_{j\in\D^{\mathrm{pair}}_t}\frac{\omega_j}{q_j}\,\mathrm{BCE}\big(\chi_j,h_\phi(s_j)\big),\\
\chi_j&=\mathbb{1}\big[|y_j-\Ch^{A}_j|<|y_j-\Ch^{T}_j|\big],\\
\omega_j&=\big||y_j-\Ch^{A}_j|-|y_j-\Ch^{T}_j|\big|,
\end{aligned}
\end{equation}
where $\D^{\mathrm{pair}}_t\subseteq\D_t$ contains matured samples on which both predictions are valid, $q_j$ is the planned probability that sample $j$ receives an Agent prediction (\S\ref{sec:route}), and samples with equal errors have zero weight. The judge uses only information available before any LLM call, and its thresholds are set on validation data.

CASH scores accounts by an ex-ante call value $\varphi(s)$, the predicted reduction in absolute error from Judge-based source selection relative to Tool. The regressor uses pre-call information, including $p_A$ and the number of Tool-licensed levels (training details in Appendix~\ref{app:impl}). The prediction agent is called for the highest-scoring accounts within the routine quota $\rho_q$,
\begin{equation}
\label{eq:gate}
M(s_i)=\mathbb{1}\big[i\in\operatorname{Top}_{\lfloor\rho_q|\mathcal I|\rfloor}(\varphi)\big],
\end{equation}
where $\operatorname{Top}_b(\varphi)$ selects the $b$ highest-scoring accounts, breaking threshold ties to keep exactly $b$ accounts. Accounts with $M(s)=0$ enter allocation with the numerical curve and its licensed levels, or take \textsc{Fallback} when $\mathcal L^{T}(s)=\emptyset$. To keep Agent errors observable on these accounts, CASH calls the prediction agent on a random audit sample of them with probability $\nu$ and uses the results only for capability learning, so the planned probability of an Agent prediction is $q_i=1$ if $M(s_i)=1$ and $q_i=\nu$ otherwise.

Accounts with $M(s)=1$ call the prediction agent for the current decision, after which the plugin computes $\mathcal L^{A}(s)$; if the call fails or returns an invalid curve, $\mathcal L^{A}(s)=\emptyset$. The option set $\mathcal O(s)$ contains \textsc{Fallback}, which gives no discount, \textsc{Tool} if $\mathcal L^{T}(s)\ne\emptyset$ and \textsc{Agent} if $\mathcal L^{A}(s)\ne\emptyset$. Licenses and $p_A$ determine the default route,
\begin{equation}
o_0(s)=\begin{cases}
\text{\textsc{Agent}}, & \mathcal L^{A}\ne\emptyset\land\big(\mathcal L^{T}=\emptyset\lor p_A>\tau_p\big),\\
\text{\textsc{Tool}}, & \mathcal L^{T}\ne\emptyset\text{ and not the above},\\
\text{\textsc{Fallback}}, & \text{otherwise}.
\end{cases}
\end{equation}
When the Agent prediction fails, the route falls back to \textsc{Tool} and then \textsc{Fallback}. Among routine-call accounts with at least one Agent-licensed level, those with ambiguous preferences, strongly diverging predictions on common licensed levels, or disjoint licenses are reviewed (conditions in Appendix~\ref{app:route}). For them, CASH assembles both curves with their errors, support, licensed levels and $p_A$ into a capability report and calls the orchestrating agent, an LLM separate from the prediction agent. The orchestrating agent chooses within $\mathcal O(s)$. It can switch between licensed sources or withhold the discount on business grounds, for example when operating information signals an imminent budget change, and a failed call or an output outside $\mathcal O(s)$ reverts to $o_0(s)$.

\subsection{Shared-Budget Allocation}
\label{sec:alloc}

Routing determines each account's curve and executable levels, with $(\tilde C_i,\K_i)=(\Ch^m,\mathcal L^m(s_i))$ for source $m$. Let $\mathcal I_{\mathrm{opt}}$ be the accounts that take part in optimization. Accounts on \textsc{Fallback} receive no discount and stay outside it; their expected revenue at $k=1$ enters the constraints as an external fixed term $C_{\mathrm{ext}}$, together with the fixed exploration configurations of \S\ref{sec:evolve}, whose subsidy forms $S_{\mathrm{ext}}$. Let $x_{ik}\in\{0,1\}$ indicate that account $i\in\mathcal I_{\mathrm{opt}}$ takes level $k$, $\tilde S_i(k)=\frac{1-k}{k}\tilde C_i(k)$, $\tilde C_{\mathrm{tot}}(x)=C_{\mathrm{ext}}+\sum_{i,k}x_{ik}\tilde C_i(k)$ and $\tilde S_{\mathrm{tot}}(x)=S_{\mathrm{ext}}+\sum_{i,k}x_{ik}\tilde S_i(k)$. With positive total discounted revenue, Eq.~\eqref{eq:asa} becomes a multiple-choice knapsack:
\begin{equation}
\label{eq:mckp}
\begin{aligned}
\max_x\quad &\sum_{i,k}x_{ik}\tilde C_i(k)\\
\text{s.t.}\quad &\tilde S_{\mathrm{tot}}(x)\ge\rho_{\min}\tilde C_{\mathrm{tot}}(x),\\
&\tilde S_{\mathrm{tot}}(x)\le\rho_{\max}\tilde C_{\mathrm{tot}}(x),\\
&\sum_{k\in\K_i}x_{ik}=1,\quad x_{ik}\in\{0,1\}.
\end{aligned}
\end{equation}

Eq.~\eqref{eq:mckp} has two cross-account constraints. Since $\rho_{\min}<\rho_{\max}$ and $\tilde C_{\mathrm{tot}}>0$, they cannot both be tight, so its LP relaxation has a basic optimal solution with at most one fractional account (Appendix~\ref{app:lp}). We search over dual prices $\mu$, round the fractional account and attempt repair through single-account swaps or \textsc{Fallback} switches (Appendix~\ref{app:route}).

The exploration actions of \S\ref{sec:evolve} form a separate channel that needs no license; after the business-safety check they enter Eq.~\eqref{eq:mckp} as fixed configurations under the same budget.

\subsection{Evidence Acquisition and Capability Updates}
\label{sec:evolve}

Matured feedback updates regional admission, which changes subsequent licenses through Eq.~\eqref{eq:license}. The \emph{Expected Policy Coverage Gain} (EPCG) allocates a limited feedback quota to regions where new evidence is expected to expand valuable Tool coverage.

We partition states and actions into regions, with $r(s,k)$ the region of $(s,k)$ (binning in Appendix~\ref{app:impl}). For matured feedback $j$ on which source $m$ has a valid prediction, we record the exceedance $z^m_j=\mathbb{1}[|y_j-\Ch^m_j|>\eta_e(s_j)]$, with a tolerance $\eta_e(s)$ proportional to the account's historical revenue. The numerical model has predictions on all samples; the prediction agent has predictions only on accounts selected by the call quota and on successfully called audit samples. Regional evidence uses only feedback in $\D^H_t$, weighted by $w^m_j(t)=\delta^{\,t-t_j}/q^m_j$ with $q^{T}_j=1$, $q^{A}_j=q_j$ and $\delta\in(0,1]$. Here $t_j$ is execution time and $t_j+L$ is maturity time; $\D^H_t$ uses maturity times. The decay favors newer evidence and the inverse probability corrects the audit sampling.

Repeated observations of the same account are highly correlated, so we treat the account as the independent unit. Let $\mathcal J^m_{r,i}$ be the evidence of account $i$ in region $r$ within the window. With account-level weight $W^m_{r,i}=\sum_{j\in\mathcal J^m_{r,i}}w^m_j$ and exceedance rate $\hat\psi^m_{r,i}=\sum_{j}w^m_jz^m_j/W^m_{r,i}$, for non-empty regions, the weighted exceedance rate and account-level Kish effective sample size are
\begin{equation}
\hat\psi^m_r=\frac{\sum_iW^m_{r,i}\hat\psi^m_{r,i}}{\sum_iW^m_{r,i}},\qquad n^{\mathrm{eff},m}_{r}=\frac{\big(\sum_iW^m_{r,i}\big)^2}{\sum_i\big(W^m_{r,i}\big)^2}.
\end{equation}
$n^{\mathrm{eff},m}_r$ is at most the number of accounts with evidence and shrinks as weight concentrates on few of them. The approximate posterior of the regional exceedance probability is
\begin{equation}
\label{eq:beta}
\psi^m_r\mid\D_t\ \dot\sim\ \mathrm{Beta}\big(n^{\mathrm{eff},m}_r\hat\psi^m_r+\tfrac12,\ n^{\mathrm{eff},m}_r(1-\hat\psi^m_r)+\tfrac12\big),
\end{equation}
a weighted approximation with the effective sample size as pseudo-count (Appendix~\ref{app:kish}). Empty regions use $n^{\mathrm{eff},m}_r=0$ and the $\mathrm{Beta}(\tfrac12,\tfrac12)$ prior directly, with $A^m_r=0$. For non-empty regions, admission requires an upper credible bound within tolerance:
\begin{equation}
\label{eq:admit}
A^m_r=\mathbb{1}\Big[\mathrm{UCB}_{1-\gamma}\big(\psi^m_r\big)\le\beta\Big],
\end{equation}
where $\mathrm{UCB}_{1-\gamma}$ is the $1-\gamma$ quantile of the posterior in Eq.~\eqref{eq:beta}. Admission is recomputed every cycle as evidence arrives or expires, so a region can be opened, kept or revoked. After retraining, Tool licenses are recomputed using the updated model's errors on held-out feedback. $A^m_r$ thus changes subsequent licenses, call selection and the levels available in allocation.

CASH aims to hand more repetitive decisions to the low-cost numerical model, so the objective of exploration is its value-weighted account and level coverage on reference traffic,
\begin{equation}
\mathrm{PC}=\frac{\sum_{s\in\mathcal S_{\mathrm{ref}}}\sum_{k\in\K}V(s)\,\mathbb{1}\big[k\in\mathcal L^{T}(s)\big]}{|\K|\sum_{s\in\mathcal S_{\mathrm{ref}}}V(s)}\in[0,1],
\end{equation}
where the reference traffic $\mathcal S_{\mathrm{ref}}$ is built from account states of the most recent cycle known at decision time and $V(s)$ is an account's value weight, its expected discounted revenue known at decision time. EPCG estimates the change in this coverage through regional admission $A^{T}_r$, with the current model and the other license conditions fixed. Model retraining is a subsequent update.

Because feedback matures with delay $L$, EPCG compares admission at $t+L$ with and without exploration. Without it, existing evidence ages under the window and decay. With it, each of $B$ simulations draws $\psi^{T,(b)}_r$ from Eq.~\eqref{eq:beta}, adds $n$ simulated outcomes $z\sim\mathrm{Bernoulli}(\psi^{T,(b)}_r)$ from distinct accounts of a frozen, randomly assigned regional candidate pool $\mathcal P_r$, and recomputes admission. Each simulated outcome enters the account-level aggregation as one observation of its account, using the regional exceedance rate as an approximation for the candidate pool (Appendix~\ref{app:route}). Then
\begin{equation}
\label{eq:epcg}
\mathrm{EPCG}_r(n)=V_r\Big[\frac1B\sum_{b=1}^{B}A^{T,(b,n)}_r(t+L)-A^{T,0}_r(t+L)\Big],
\end{equation}
where $V_r$ is the normalized weight of account and level pairs in region $r$ that satisfy every condition of Eq.~\eqref{eq:license} except admission, using the same weights and denominator as $\mathrm{PC}$. EPCG is negative when new feedback is likely to reveal larger errors and revoke admission. It targets evidence that confirms where the current model can be trusted, and regions that also fail the support or error conditions are left to natural feedback and retraining.

Admission is a threshold decision, so EPCG changes in steps with $n$, not necessarily monotonically. Given feedback quota $N_{\mathrm{fb}}$, we optimize over discrete packet sizes $\mathcal N$ with $0\in\mathcal N$:
\begin{equation}
\label{eq:quota}
\begin{aligned}
\max_{\{n_r\}}\quad &\sum_r\mathrm{EPCG}_r(n_r)\\
\text{s.t.}\quad &\sum_r n_r\le N_{\mathrm{fb}},\\
&n_r\in\mathcal N,\quad n_r\le|\mathcal P_r|.
\end{aligned}
\end{equation}
Since the regional pools are disjoint, Eq.~\eqref{eq:quota} is a multiple-choice knapsack solved exactly by dynamic programming. Explorations are executed one cycle at a time and are cut from the lowest EPCG per unit if Eq.~\eqref{eq:mckp} becomes infeasible.

\section{Experiments}
\label{sec:exp}

We evaluate CASH through six research questions.
\begin{list}{}{\setlength{\leftmargin}{0pt}\setlength{\rightmargin}{0pt}\setlength{\itemindent}{0pt}\setlength{\listparindent}{0pt}\setlength{\labelwidth}{0pt}\setlength{\labelsep}{0pt}}
\item \textbullet\enspace \textbf{RQ1: Online impact.} How does CASH perform relative to the incumbent policy online?
\item \textbullet\enspace \textbf{RQ2: Prediction accuracy.} How does CASH compare with individual estimators and alternative combinations?
\item \textbullet\enspace \textbf{RQ3: Capability reliability.} Are the estimators complementary, and do capability estimates identify reliable predictions and improve review by the orchestrating agent?
\item \textbullet\enspace \textbf{RQ4: Call selection and licenses.} What are the effects of call selection and capability licenses on accuracy and inference cost?
\item \textbullet\enspace \textbf{RQ5: Budget coupling.} How do local estimation errors propagate through the shared budget?
\item \textbullet\enspace \textbf{RQ6: Capability evolution.} Does EPCG expand reliable coverage and improve subsequent predictions?
\end{list}

\subsection{Setup}
\label{sec:setup}

\subsubsection{Online experiment}
CASH, Tool and Control serve 60\%, 20\% and 20\% of platform traffic, respectively, with group assignment based on hashed account IDs. A pre-experiment A/A test verified that the groups were comparable in revenue per account, ADVV and subsidy rate. Control retains the incumbent tier-based policy guided by operator experience. Tool combines the numerical model and budget allocator. CASH's full harness ran in production for one month from September 14 to October 14, 2026.

\subsubsection{Offline data and evaluation cohort}
We construct a dataset of 663,312 account-days from real advertising logs. Each sample contains the account history available before the target period, the executed discount factor and the realized discounted revenue in that period. The chronological training, validation and test splits contain 554,724, 54,851 and 53,737 account-days. Validation data serve model selection, error calibration and threshold setting. The 16 candidate discount factors range from 0.50 to 0.80, in addition to the no-discount action $k=1$. For offline method comparisons involving LLMs, we use a common subset of 20,143 account-days from 11,314 accounts across two test periods, evaluating all methods at historically executed subsidy levels.

\subsubsection{Compared methods}
Tool uses the numerical response model alone and, online, pairs it with the budget allocator. Agent uses the LLM-based prediction agent alone. Agent Routing lets an LLM select between the two predictions, following the LLM-as-a-judge paradigm~\cite{zheng2023judging}. Judge Routing uses a learned source selector without capability constraints, following RouteLLM~\cite{ong2024routellm}. CASH combines selective LLM calls with capability-constrained routing and allocation, and CASH (all calls) obtains both predictions and invokes the orchestrating agent for every account. Implementation details are in Appendix~\ref{app:impl}.

\subsubsection{Metrics and protocol}
Revenue, ADVV and LTV are reported per account using all advertisers in each group as the denominator. Revenue is post-discount platform revenue, ADVV is the advertiser value generated by acquired traffic, and LTV is cumulative revenue over 7 and 14 days. Target-achieving share is the fraction of revenue from accounts meeting their optimization targets, ineffective share is the fraction of revenue without conversion, and retention is the proportion of accounts active on days 7 and 14. Online results use Control as the reference unless otherwise stated. Gains in Revenue, ADVV and LTV are relative changes, while changes in revenue shares and retention are reported in percentage points. Offline, MAE and RMSE measure revenue prediction error at executed levels. CASH uses the Tool prediction for scoring when it takes \textsc{Fallback} or the executed level is outside the selected source's license. A severe exceedance is an absolute error above twice the account's pre-prediction historical revenue scale. Cost counts every underlying LLM request, including prediction, audit and orchestration calls. Offline paired comparisons use a 1,000-replicate account-clustered bootstrap for 95\% confidence intervals, RQ6 applies Holm correction for multiple comparisons, and Appendix~\ref{app:online_uncertainty} details the daily-aggregate analysis of online temporal uncertainty.

\subsection{RQ1: Online Impact}
\label{sec:rq1}

\subsubsection{Overall performance}
CASH leads in revenue per account, ADVV, LTV and retention, as shown in Table~\ref{tab:online}. At essentially the same realized subsidy rate as Control, CASH raises revenue per account by 15.35\% and ADVV by 19.86\%, increases the target-achieving share by 2.29 percentage points and lowers the ineffective share by 1.35 points. Seven- and 14-day LTV gains are 15.98\% and 15.24\%, while day-14 retention gains 1.57 points.

\begin{table*}[t]
\caption{Online performance of Tool and CASH relative to Control across revenue, advertiser value, revenue quality, LTV and retention. Revenue, ADVV and LTV are reported as relative gains per account; revenue shares and retention are reported as changes in percentage points.}
\label{tab:online}
\centering
\begin{tabular*}{\textwidth}{@{\extracolsep{\fill}}lcccccccc@{}}
\toprule
& \multicolumn{2}{c}{Revenue and ADVV} & \multicolumn{2}{c}{Revenue quality} & \multicolumn{2}{c}{LTV per account} & \multicolumn{2}{c}{Retention} \\
\cmidrule(lr){2-3}\cmidrule(lr){4-5}\cmidrule(lr){6-7}\cmidrule(lr){8-9}
Method & Revenue $\uparrow$ & ADVV $\uparrow$ & Target-achieving $\uparrow$ & Ineffective $\downarrow$ & 7-day $\uparrow$ & 14-day $\uparrow$ & Day 7 $\uparrow$ & Day 14 $\uparrow$ \\
\midrule
Tool & +8.50\% & +14.32\% & $-$1.42\,pp & $-$1.78\,pp & +12.51\% & +11.50\% & +1.04\,pp & +1.40\,pp \\
CASH & +15.35\% & +19.86\% & +2.29\,pp & $-$1.35\,pp & +15.98\% & +15.24\% & +1.15\,pp & +1.57\,pp \\
\bottomrule
\end{tabular*}

\end{table*}

The Tool group provides a numerical-model allocation baseline for assessing the additional gains of CASH. Automated allocation with the numerical model alone already raises ADVV by 14.32\% over Control. Over Tool, CASH gains 6.31\% in revenue per account, 3.08\% in seven-day LTV and 3.72 points in target-achieving share, while ineffective share rises by 0.43 points.

\begin{figure}[t]
\centering
\includegraphics[width=\columnwidth]{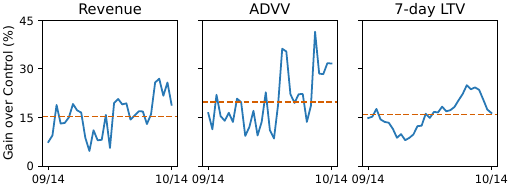}
\caption{Daily per-account gains of CASH over Control. Dashed lines indicate pooled gains.}
\Description{Daily relative gains in Revenue, ADVV, and seven-day LTV are positive on all 31 anchor dates. Dashed lines indicate the corresponding pooled gains.}
\label{fig:online_daily}
\end{figure}

\subsubsection{Temporal stability}
The gains hold throughout the test period from September 14 to October 14, 2026, as shown in Figure~\ref{fig:online_daily}. Revenue, ADVV and seven-day LTV per account are higher under CASH than under Control on each of the 31 days, and removing any single day keeps the pooled gain between 14.98\% and 15.71\% for revenue and between 19.17\% and 20.25\% for ADVV. A paired seven-day moving-block bootstrap over daily aggregates with 50,000 replicates yields approximate 95\% temporal intervals of $[12.05\%,17.77\%]$ for revenue, $[15.05\%,23.18\%]$ for ADVV and $[11.86\%,19.63\%]$ for seven-day LTV, all above zero. Appendix~\ref{app:online_uncertainty} reports positive intervals for alternative block lengths.

\subsection{RQ2: Offline Prediction Accuracy}
\label{sec:rq2}

\begin{figure}[t]
\centering
\includegraphics[width=\columnwidth]{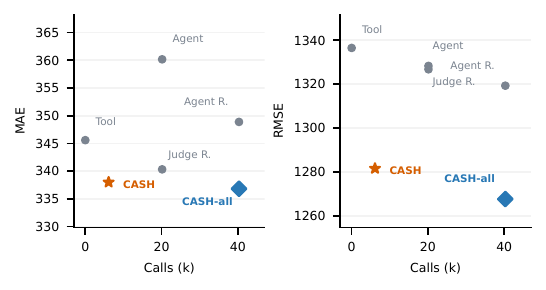}
\caption{Prediction error versus LLM calls. Lower and further left is better. R. denotes Routing; CASH-all denotes CASH (all calls).}
\Description{MAE and RMSE versus LLM calls for six methods. CASH uses 6,119 calls, while CASH with all calls uses 40,286 and achieves the lowest errors.}
\label{fig:accuracy_cost}
\end{figure}

As shown in Figure~\ref{fig:accuracy_cost} and Table~\ref{tab:prediction}, neither individual source leads on both error metrics. Agent has a higher MAE than Tool, 360.17 against 345.58, but a lower RMSE, 1328.32 against 1336.51. RQ3 examines their sample-level complementarity. Simple routing gives limited improvements: Agent Routing leaves MAE above Tool, while Judge Routing lowers MAE with a smaller RMSE gain.

CASH lowers MAE by 2.20\% and RMSE by 4.11\% relative to Tool, with a 95\% confidence interval of $[-12.29,-3.37]$ for the paired MAE difference. Its larger relative reduction in RMSE is consistent with the tail-error reduction observed in the license ablation of RQ4. CASH uses 6,119 calls, fewer than any baseline that calls the LLM for every account.

\subsection{RQ3: Complementarity and Capability Reliability}
\label{sec:rq3}

\begin{figure}[t]
\centering
\begin{subfigure}[t]{0.48\columnwidth}
\centering
\includegraphics[width=\linewidth]{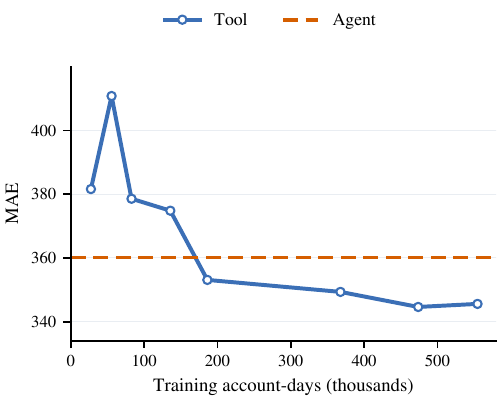}
\caption{MAE versus training-set size.}
\label{fig:capcurve}
\end{subfigure}\hfill
\begin{subfigure}[t]{0.48\columnwidth}
\centering
\includegraphics[width=\linewidth]{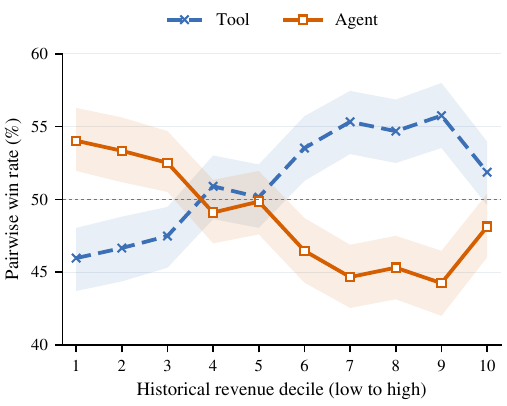}
\caption{Win rates by revenue decile.}
\label{fig:winrate}
\end{subfigure}
\caption{Estimator complementarity across training-set sizes and revenue deciles. Shaded bands in (b) show 95\% CIs.}
\Description{Panel a compares Tool MAE across training-set sizes with a fixed Agent. Panel b shows Tool and Agent win rates across revenue deciles, with confidence intervals.}
\label{fig:complementarity}
\end{figure}

As the training set grows, Tool overtakes the fixed Agent, as shown in Figure~\ref{fig:capcurve}. With limited training data, Tool performs worse than Agent and fluctuates across training sizes, with an MAE of about 411 at 56,000 account-days. As training data accumulate, its MAE falls below Agent's 360.17 between 136,000 and 186,000 account-days and reaches 345.58 at about 555,000. The final predictors remain complementary, since Agent is more accurate on 48.77\% of account-days although Tool has the lower MAE. Agent wins more often on small accounts and Tool on large ones, as shown in Figure~\ref{fig:winrate}, yet win rates remain between 44\% and 56\% across revenue deciles. Appendix~\ref{app:additional} reports prediction errors and Agent win rates for accounts with low, medium and high Tool support.

\begin{table}[t]
\caption{Reliability signals at matched admission. Exceed. denotes severe exceedance.}
\label{tab:reliability}
\centering\small
\begin{tabular*}{\columnwidth}{@{\extracolsep{\fill}}llcc@{}}
\toprule
Source & Admission signal & MAE$\downarrow$ & Exceed.$\downarrow$ \\
\midrule
Tool & Support & 371.95 & 9.61\% \\
 & Source disagreement & 272.43 & 7.53\% \\
 & Local calibration & \textbf{230.53} & \textbf{4.94\%} \\
\midrule
Agent & Support & 378.73 & 10.85\% \\
 & Source disagreement & 270.26 & 8.13\% \\
 & Local calibration & \textbf{250.83} & \textbf{6.12\%} \\
\bottomrule
\end{tabular*}
\end{table}

Capability estimates learned from realized errors separate risky predictions. Under the deployed licenses, the severe-exceedance rate is 6.29\% inside and 15.08\% outside the Tool licenses, and 5.67\% and 14.97\% for Agent. Table~\ref{tab:reliability} compares admission signals at approximately 60\% admission, with matched counts within account-size buckets. Among the compared signals, local calibration gives the lowest MAE and exceedance rate for both sources. The admitted Tool MAE is 38.0\% below support-based and 15.4\% below disagreement-based admission; the admitted Agent MAE is 7.2\% below disagreement-based admission. Compared with a global constant error bound, local calibration reduces the median bound by 35.8\% for Tool and 30.3\% for Agent, with empirical coverage of approximately 94\% against a nominal 95\%. Exposing this information also constrains review by the orchestrating agent, as shown in Table~\ref{tab:review} in Appendix~\ref{app:additional}. With the license list, out-of-license choices drop from 66 to zero, and the full capability report lowers MAE on the difficult review subset from 479.60 to 466.96.

\subsection{RQ4: Call Selection and Licenses}
\label{sec:rq4}

Call selection retains most of the gain of CASH (all calls) at a small fraction of the cost. CASH (all calls) lowers MAE from 345.58 for Tool to 336.80, and CASH reaches 337.99, recovering 86\% of this reduction with 6,119 calls, 15\% of the 40,286 calls of the full configuration. Relative to the Agent baseline with one call per account-day, CASH uses 69.6\% fewer calls and 68.0\% fewer tokens. Raising the routine quota to 40\% and 60\% gives an MAE of 338.22 and 338.34, respectively, as reported in Appendix~\ref{app:additional}.

Licenses mainly control tail error. We remove only the license constraint on source selection and keep quota selection, predictions and review responses fixed. MAE stays nearly unchanged at 337.92 against 337.99, while RMSE rises by 8.78 from 1281.63 to 1290.41, with a 95\% confidence interval of $[0.73,24.18]$ for the increase. The increase in RMSE with little change in MAE indicates that licenses primarily limit large prediction errors.

\subsection{RQ5: Budget Coupling}
\label{sec:rq5}

Motivated by the budget-coupling mechanism in Proposition~\ref{prop:scale}, we stress-test how single-account estimation errors propagate through the shared budget. Each of 672 scenarios rescales the prediction of one account on either its licensed or its unlicensed levels, re-solves Eq.~\eqref{eq:mckp} under the deployed subsidy-rate interval with and without licenses, and counts the level changes of other accounts. Strict switches exclude changes due solely to ties between equally valued allocations.

\begin{table}[t]
\caption{Cross-account effects of single-account perturbations over 672 scenarios. Strict switches exclude changes due solely to ties between equally valued allocations.}
\label{tab:coupling}
\small
\begin{tabular*}{\columnwidth}{@{\extracolsep{\fill}}llccc@{}}
\toprule
\multicolumn{3}{c}{} & \multicolumn{2}{c}{Strict switches$\downarrow$} \\
\cmidrule(lr){4-5}
Perturbed levels & Licenses & Changes$\downarrow$ & Mean & Max \\
\midrule
Unlicensed & Off & 0.51 & 0.010 & 1 \\
 & On & \textbf{0} & \textbf{0} & \textbf{0} \\
\midrule
Licensed & Off & 1.09 & 0.680 & 33 \\
 & On & \textbf{0.90} & \textbf{0.480} & \textbf{29} \\
\bottomrule
\end{tabular*}
\end{table}

Licenses block propagation from unlicensed levels, as shown in Table~\ref{tab:coupling}. Without licenses, perturbations on these levels cause an average of 0.510 level changes and 0.010 strict switches in other accounts, indicating that most changes occur between equally valued allocations. With licenses, these levels are excluded and propagation falls to zero. Perturbations on licensed levels still affect other accounts. Without licenses, they cause 1.090 changes on average and up to 33 strict switches, and with licenses, mean changes fall to 0.900 and mean strict switches from 0.680 to 0.480.

\subsection{RQ6: Capability Evolution}
\label{sec:rq6}

To evaluate coverage expansion and subsequent prediction gains, we withhold weekend samples, accounts with recent revenue spikes, or large high-volatility accounts from initial Tool training, and replay acquisition on held-out logs, where each method selects samples under the same quota and observes their logged outcomes. With Tool frozen and only admission updated, the three configurations in Table~\ref{tab:coverage} yield value-weighted licensed coverage gains between 2.06 and 2.75 points for EPCG and between $-0.37$ and $+0.35$ points for random acquisition. Across all 15 gap--quota combinations, EPCG leads in every combination and wins 44 of 45 seed-level comparisons with one tie. Its mean advantage across the remaining combinations is between 1.78 and 2.17 points when any single gap is left out. Samples in newly admitted regions retain a severe-exceedance rate between 4.57\% and 5.79\% in later evaluation periods, close to that of licensed predictions in RQ3. Predicted EPCG correlates with realized coverage gains, with Spearman correlations of 0.578 across selected regions and 0.778 across complete acquisition plans. On the high-volatility gap, we retrain Tool on the acquired feedback and evaluate it on 27,613 later account-days. With 480, 960 and 1,920 feedback samples, EPCG lowers RMSE relative to random acquisition by 4.1\%, 3.0\% and 1.5\% while MAE stays similar, as shown in Table~\ref{tab:future} in Appendix~\ref{app:additional}. An agent-guided variant reserves half of the quota for strata proposed by the agent from a diagnostic report. It has the lowest MAE, but no significant Holm-adjusted advantage.

\section{Related Work}

\subsection{Budget-Constrained Incentive Allocation}
Incentive allocation requires estimating heterogeneous responses and coordinating decisions under a shared budget. Uplift models estimate treatment effects~\cite{zhao2017uplift,zhong2022descn}, while budget-aware allocation methods connect response estimation to constrained optimization~\cite{zhao2019unified,ai2022lbcf}. More recent approaches tighten this connection through end-to-end optimization or decision-oriented causal learning~\cite{sun2024e3ir,zhou2023dhcl}. In advertising, related work studies budget-constrained bidding~\cite{wu2018budget} and coupon mechanisms for advertiser participation and retention~\cite{shen2021coupon,xing2025autocoupon}. CASH addresses a complementary issue: which response estimates should be allowed to influence allocation when historical interventions provide uneven support. Its account- and level-specific licenses restrict the estimates used by the shared-budget optimizer, while targeted feedback updates these restrictions over time.

\subsection{Capability-Aware LLM Agents}
LLM agents combine reasoning with external tools~\cite{yao2023react,schick2023toolformer}, including for bidding and advertising budget allocation~\cite{cai2025rtbagent,song2026dara}. Their use as numerical estimators~\cite{gruver2023llmtime} also raises questions about calibration and the reliability of confidence signals~\cite{epstein2025fermieval,qin2026quantsight,xiao2025consistency,tao2025revisiting}. Learning to defer and selective prediction study when to rely on a predictor or seek assistance~\cite{mozannar2020consistent,charusaie2022sample,geifman2017selective}; LLM routing and tool-use policies balance predictive quality against inference cost~\cite{chen2024frugalgpt,ding2024hybrid,ong2024routellm,qian2025smart}. CASH makes a finer distinction between selecting a source and determining the actions on which that source may be used. Outcome-based licenses constrain both the orchestrating agent's source choice and the subsequent allocation across accounts. This design also relates to uncertainty-aware decision-making and active evidence acquisition. Conformal prediction calibrates uncertainty, including for downstream decisions~\cite{lei2018distribution,cortesgomez2025utility}, while conservative policy learning and safe exploration restrict actions under limited evidence~\cite{laroche2019spibb,kumar2020cql,sui2015safe,wu2016conservative}. Level-set estimation explicitly targets growth of a certified region~\cite{zanette2018robust}; value-of-information and budgeted exploration study how to spend limited opportunities to learn~\cite{frazier2008knowledge,badanidiyuru2013bwk}. EPCG evaluates feedback by its expected effect on future value-weighted reliable coverage of the numerical model, measured through changes in regional admission. Acquired evidence updates the model and regional admission to guide subsequent allocation.

\section{Conclusion}
We studied Advertiser Subsidy Allocation, which turns experience-driven subsidy configuration into automated account-level decisions under a platform-wide subsidy-rate constraint. The central challenge is to combine estimators whose reliability varies across accounts and subsidy levels while controlling how estimation errors affect shared-budget allocation. CASH combines an LLM-based prediction agent with a numerical response model, learns per-level capability licenses for both sources from realized outcomes, and uses these licenses to constrain source selection, review by the orchestrating agent and shared-budget allocation. EPCG prioritizes feedback by its expected contribution to reliable Tool coverage. Offline evaluations show that licenses reduce tail prediction errors and block propagation from unlicensed estimates, while EPCG expands reliable coverage more effectively than random acquisition. In a one-month online experiment with CASH serving 60\% of platform traffic, CASH raises revenue per account by 15.35\% and ADVV by 19.86\% over Control at an essentially unchanged subsidy rate. It further increases revenue per account by 6.31\% over automated allocation with the numerical model alone. These results demonstrate the practical value of combining operational knowledge with data-driven prediction within outcome-grounded capability boundaries for fine-grained advertiser subsidy allocation, improving platform revenue and advertiser outcomes under a shared budget.

\FloatBarrier
\bibliographystyle{ACM-Reference-Format}
\bibliography{sample-base}

\appendix
\numberwithin{equation}{section}

\section{LP Relaxation: Fractional Accounts}
\label{app:lp}
Let $N_{\mathrm{opt}}=|\mathcal I_{\mathrm{opt}}|$. The constraints of the LP relaxation consist of $N_{\mathrm{opt}}$ account-choice equalities and two budget inequalities; after adding slack variables, every basic feasible solution has $N_{\mathrm{opt}}+2$ basic variables. Because $\rho_{\min}<\rho_{\max}$ and the feasible solutions considered satisfy $\tilde C_{\mathrm{tot}}(x)>0$, the two budget constraints cannot both be tight (otherwise $(\rho_{\max}-\rho_{\min})\tilde C_{\mathrm{tot}}(x)=0$), so at least one slack variable is positive and basic, and at most $N_{\mathrm{opt}}+1$ basic variables come from $x$. Each account has at least one $x_{ik}>0$, so at most one account has two positive components, that is, at most one account is fractional. If neither budget constraint is tight, all accounts are integral. The property holds for basic optimal solutions; an interior-point solution requires crossover. It bounds the number of fractional accounts, not the objective loss after rounding or the feasibility of the integer problem.

\section{Implementation Details}
\label{app:impl}
Candidate discount factors are $0.50,0.52,\ldots,0.80$. The target subsidy rate is 68\%, with an allowed interval of $[67.9\%,68.1\%]$. Qwen3-Next-80B-A3B-Instruct estimates Agent response curves. DeepSeek-V4-Flash orchestrates the harness, reads capability reports, and performs review. CASH uses a routine quota of 20\% and an audit rate of 5\%. In CASH (all calls), each evaluated account-day incurs one estimation request and one orchestration request. The judge is a gradient-boosted tree and requires no LLM call. In the license ablation of RQ4, the reused review responses were generated with license information in the prompt. Error intervals use a nominal 95\% coverage level. Regional admission uses a 95\% posterior upper bound and a 10\% tolerated exceedance probability. The exploration settings are $\{0,5,10,20,40\}$, and HiGHS independently verifies the allocation. Account scale is the mean discounted revenue over the previous 7 cycles, and its $\log(1+\cdot)$ octiles on development data define 8 scale buckets $b(s)$ that stay frozen during evaluation. Regions cross these buckets with the 16 levels, giving 128 regions, and historical off-grid actions are mapped to the nearest level. The support kernels of Eq.~\eqref{eq:support} are $\kappa_s(s,s')=\mathbb{1}[b(s)=b(s')]$ and $\kappa_k(k_j,k_{j'})=\frac13\mathbb{1}[|j-j'|\le1]$ over level indices, so support comes from the same scale bucket and from the same or an adjacent level. Regional evidence is kept for at most $H=14$ cycles with decay $\delta=0.95$. Matched-admission analysis operates at 60\% coverage.

In the quota experiments, the call-value regressor targets $g_j=|y_j-\Ch^T_j|-|y_j-\Ch^{J,\mathrm{OOF}}_j|$, where $\Ch^{J,\mathrm{OOF}}_j$ is the prediction selected by a cross-fitted Judge. Signed gains are symmetrically clipped at the training-set 99th-percentile threshold and fitted with squared loss, measuring prediction-error reduction before orchestration.

\section{Routing, Allocation and Exploration Details}
\label{app:route}
With an Agent-licensed level, ambiguous preferences or strongly diverging predictions trigger review,
\begin{equation}
\begin{aligned}
\mathrm{Review}(s)={}&M(s)\,\mathbb{1}\big[\mathcal L^{A}(s)\ne\emptyset\big]\\
&\cdot\mathbb{1}\big[|p_A(s)-\tau_p|<\eta_p\lor D(s)>\eta_D\big],
\end{aligned}
\end{equation}
where
\begin{equation}
D(s)=\begin{cases}
\displaystyle\max_{k\in\mathcal L^{T}\cap\mathcal L^{A}}\frac{|\Ch^{A}_k-\Ch^{T}_k|}{\max(\Ch^{T}_k,c_0)}, & \text{overlapping},\\[6pt]
+\infty, & \text{non-empty, disjoint},\\[2pt]
0, & \text{otherwise},
\end{cases}
\end{equation}
with $\Ch^m_k=\Ch^m(s,k)$. For two non-empty disjoint licenses, $D(s)=+\infty$ marks a license conflict and also triggers review. For reviewed accounts, CASH assembles both curves with their errors, support, licensed levels and $p_A$ into a capability report $\mathrm{Rep}(s)$ and calls the orchestrating agent, whose choice $\pi_{\mathrm{LLM}}(\mathrm{Rep}(s),\xi_s)$ lies in $\mathcal O(s)$:
\begin{equation}
\label{eq:agent}
o(s)=\begin{cases}
\pi_{\mathrm{LLM}}\big(\mathrm{Rep}(s),\xi_s\big),
&\begin{gathered}\mathrm{Review}(s),\\\pi_{\mathrm{LLM}}(\mathrm{Rep}(s),\xi_s)\in\mathcal O(s),\end{gathered}\\[3pt]
o_0(s), & \text{otherwise}.
\end{cases}
\end{equation}

Eq.~\eqref{eq:mckp} has only two cross-account budget constraints. Because $\rho_{\min}<\rho_{\max}$ and $\tilde C_{\mathrm{tot}}>0$, the two cannot be tight simultaneously, so the LP relaxation has a basic optimal solution in which at most one account is fractional (Appendix~\ref{app:lp}). We solve the LP relaxation by a search over the dual prices of the two budget constraints, round the fractional account and repair the constraints by single-account swaps; HiGHS solves the same relaxation independently to verify the result. We measure repair by the two-sided violation
\begin{equation}
\begin{aligned}
v(x)={}&\big[\rho_{\min}\tilde C_{\mathrm{tot}}(x)-\tilde S_{\mathrm{tot}}(x)\big]_+\\
&+\big[\tilde S_{\mathrm{tot}}(x)-\rho_{\max}\tilde C_{\mathrm{tot}}(x)\big]_+.
\end{aligned}
\end{equation}
If the solution is still infeasible, accounts are switched to \textsc{Fallback} in decreasing order of their contribution to the violation; a switch is accepted only if it strictly decreases $v$, and the full constraints are verified again at the end. If repair and exploration cancellation still leave an infeasible allocation, the offline allocation procedure returns failure without producing a new executable plan.

Exploration accounts are sampled from frozen regional candidate pools. The pool is first filtered by business rules, keeping accounts whose levels belong to $\K^{\mathrm{safe}}(s)$, with an agent risk review added as a variant when operating information is available; once frozen, it is randomly assigned by a fixed rule to disjoint regional pools $\mathcal P_r$, and exploration accounts of region $r$ are drawn at random from $\mathcal P_r$ and executed at a level of region $r$. The pool, exclusion reasons, draw probabilities, and cancellations and replacements before execution are all logged.

Because feedback matures with delay $L$, EPCG compares two paths at the same time $t+L$. Without exploration, existing evidence ages under the window and decay, giving admission $A^{T,0}_r(t+L)$. With exploration, $n$ simulated feedback samples are added on top of the same aging. Simulation $b$ draws $\psi^{T,(b)}_r$ from Eq.~\eqref{eq:beta}, selects $n$ accounts from $\mathcal P_r$ by the actual sampling rule, generates $z^{T,(b)}_j\sim\mathrm{Bernoulli}(\psi^{T,(b)}_r)$, merges them with weight $\delta^L$ into each account's existing evidence and recomputes admission, giving $A^{T,(b,n)}_r(t+L)$. Both paths use the same reference traffic, aging rule and fixed license conditions; EPCG then follows Eq.~\eqref{eq:epcg}.

\begin{algorithm}[h]
\caption{One decision cycle of CASH (cycle $t$)}
\label{alg:loop}
\begin{algorithmic}[1]
\State Compute and project $\Ch^{T}(s_i,\cdot)$; determine $\K^{\mathrm{safe}}(s_i)$; the plugin yields $\bar u^{T}$, $\sigma$, $p_A$ and $\mathcal L^{T}$
\State Select accounts by Eq.~\eqref{eq:gate}; audit remaining accounts with the prediction agent at probability $\nu$, for learning only
\State For $M(s_i)=0$: \textsc{Tool} if $\mathcal L^{T}(s_i)\ne\emptyset$; otherwise \textsc{Fallback}
\For{each account with $M(s_i)=1$}
  \State Obtain $\Ch^{A}(s_i,\cdot)$ from the prediction agent; project as in \S\ref{sec:est} ($\mathcal L^{A}=\emptyset$ on failure); compute $\mathcal L^{A}$, $\mathcal O(s_i)$ and $o_0(s_i)$; set $o(s_i)=o_0(s_i)$
  \State If $\mathrm{Review}(s_i)$, call the orchestrating agent and update $o(s_i)$ by Eq.~\eqref{eq:agent}
\EndFor
\State Form frozen pools $\mathcal P_r$; compute EPCG by Eq.~\eqref{eq:epcg}; allocate the quota by Eq.~\eqref{eq:quota}
\State Solve Eq.~\eqref{eq:mckp} with explorations fixed; repair by $v(x)$; cut explorations if infeasible
\State In offline allocation, return failure if infeasibility remains
\State After feedback matures, update windowed evidence, $n^{\mathrm{eff},m}_r$, $A^m_r$, calibration and the judge; retrain on schedule
\end{algorithmic}
\end{algorithm}

\section{Additional Experimental Results}
\label{app:additional}

\subsection{Temporal Uncertainty of Online Gains}
\label{app:online_uncertainty}
We resample daily aggregates from September 14 to October 14, 2026, always pairing CASH and Control on the same sampled dates. The primary analysis draws five blocks with replacement from the 25 overlapping, non-circular seven-day blocks, concatenates them, and retains the first 31 dates. Each replicate recomputes group-level ratios from the summed numerators and denominators; it does not average daily lifts. We use 50,000 replicates and the 2.5th and 97.5th percentiles. Table~\ref{tab:online_blocks} reports sensitivity to block length. These are pointwise, approximate temporal intervals from 31 daily aggregates, not account-clustered or randomization-based intervals. Trends and overlapping LTV windows limit coverage claims.

\begin{table}[!htbp]
\caption{Approximate 95\% temporal bootstrap intervals for relative gains (\%).}
\label{tab:online_blocks}
\centering\small
\begin{tabular}{@{}lccc@{}}
\toprule
Block length & Revenue & ADVV & 7-day LTV \\
\midrule
3 days & [12.79, 18.24] & [16.16, 23.61] & [13.26, 18.82] \\
5 days & [12.26, 18.31] & [15.53, 23.87] & [12.41, 19.46] \\
7 days & [12.05, 17.77] & [15.05, 23.18] & [11.86, 19.63] \\
\bottomrule
\end{tabular}
\end{table}

\subsection{Further Offline Results}

\begin{table}[!htbp]
\caption{Prediction accuracy and total LLM calls.}
\label{tab:prediction}
\centering
\small
\begin{tabular}{@{}lrrr@{}}
\toprule
Method & MAE$\downarrow$ & RMSE$\downarrow$ & Calls$\downarrow$ \\
\midrule
Tool & 345.58 & 1336.51 & 0 \\
Agent & 360.17 & 1328.32 & 20,143 \\
Agent Routing & 348.87 & 1319.32 & 40,286 \\
Judge Routing & 340.31 & 1326.79 & 20,143 \\
CASH & 337.99 & 1281.63 & 6,119 \\
CASH (all calls) & \textbf{336.80} & \textbf{1267.70} & 40,286 \\
\bottomrule
\end{tabular}
\end{table}

\subsubsection{Support tiers}
Table~\ref{tab:support} groups samples by the support of the Tool in its training data, with cut points fixed from states known during development.

\begin{table}[!htbp]
\caption{Errors and Agent win rates by Tool support.}
\label{tab:support}
\centering\small
\begin{tabular}{@{}lrrrr@{}}
\toprule
Support & Samples & Tool MAE & Agent MAE & Agent wins \\
\midrule
Low & 7,488 & 442.73 & 465.33 & 46.8\% \\
Medium & 6,614 & 330.75 & 339.91 & 53.1\% \\
High & 6,041 & 241.40 & 252.02 & 46.4\% \\
\bottomrule
\end{tabular}
\end{table}


\begin{table}[!htbp]
\caption{Value-weighted licensed coverage gains with Tool frozen, in points; means over three seeds, brackets give the seed range of the difference.}
\label{tab:coverage}
\centering\small
\setlength{\tabcolsep}{4pt}
\begin{tabular}{@{}lrrrr@{}}
\toprule
Gap & Feedback & EPCG & Random & Diff. [range] \\
\midrule
Weekend & 215 & +2.06 & $-$0.29 & +2.34 [2.03, 2.67] \\
Revenue spike & 315 & +2.74 & +0.35 & +2.39 [2.33, 2.42] \\
Volatile & 320 & +2.75 & $-$0.37 & +3.12 [2.84, 3.58] \\
\bottomrule
\end{tabular}
\end{table}

\begin{table}[!htbp]
\caption{Capability information in difficult-account review. Violations count unlicensed choices.}
\label{tab:review}
\centering\small
\begin{tabular}{@{}lrrr@{}}
\toprule
Review input & Violations$\downarrow$ & MAE$\downarrow$ & RMSE$\downarrow$ \\
\midrule
No capability information & 66 & 479.60 & 1868.40 \\
License list & 0 & 474.15 & \textbf{1797.35} \\
Full capability report & 0 & \textbf{466.96} & 1805.06 \\
\bottomrule
\end{tabular}
\end{table}

\begin{table}[!htbp]
\caption{Future Tool errors for the large, high-volatility gap after updating on acquired feedback. Mean $\pm$ standard deviation over three seeds.}
\label{tab:future}
\centering\small
\setlength{\tabcolsep}{3pt}
\begin{tabular}{@{}rlrr@{}}
\toprule
Feedback & Acquisition & MAE$\downarrow$ & RMSE$\downarrow$ \\
\midrule
480 & Random & $286.21\pm0.79$ & $1143.38\pm17.56$ \\
 & EPCG & $285.64\pm0.30$ & $\mathbf{1096.33}\pm16.59$ \\
 & Agent + EPCG & $\mathbf{284.28}\pm0.48$ & $1098.90\pm17.12$ \\
\midrule
960 & Random & $285.71\pm0.40$ & $1132.32\pm10.50$ \\
 & EPCG & $285.92\pm0.25$ & $\mathbf{1098.05}\pm15.47$ \\
 & Agent + EPCG & $\mathbf{284.68}\pm0.28$ & $1101.68\pm12.23$ \\
\midrule
1,920 & Random & $285.21\pm0.38$ & $1112.99\pm16.10$ \\
 & EPCG & $286.08\pm0.15$ & $\mathbf{1096.60}\pm14.14$ \\
 & Agent + EPCG & $\mathbf{284.33}\pm0.14$ & $1105.09\pm12.55$ \\
\bottomrule
\end{tabular}
\end{table}

\subsubsection{Routine-call quotas}
Table~\ref{tab:quota} compares prediction accuracy and call counts across routine quotas.
At the 20\% quota, the 6,119 calls consist of 4,832 estimations, including 803 audits, and 1,287 reviews. The 40\% and 60\% quotas add calls without improving MAE, and CASH (all calls) improves it only slightly.

\begin{table}[!htbp]
\caption{CASH accuracy and call counts by routine quota.}
\label{tab:quota}
\centering\small
\begin{tabular}{@{}lrr@{}}
\toprule
Routine quota & Total calls & MAE$\downarrow$ \\
\midrule
20\% (deployed) & 6,119 & 337.99 \\
40\% & 11,384 & 338.22 \\
60\% & 16,446 & 338.34 \\
CASH (all calls) & 40,286 & 336.80 \\
\bottomrule
\end{tabular}
\end{table}

\subsubsection{License conditions}
On the same common cohort, with the same account-size buckets and 12,085 admitted samples per source and method, adding support to local calibration lowers Tool MAE from 230.53 to 229.09 and Agent MAE from 250.83 to 250.37. Adding regional admission further lowers these errors to 211.92 and 220.72, respectively.

\subsubsection{Acquisition baselines}
In an independent masked replay, the average Spearman correlation between the predicted regional gain and the realized coverage gain is 0.384 for EPCG, 0.352 for value-first, 0.134 for boundary-first and 0.078 for uncertainty-first acquisition; value-first is better on some slices. On the replay of RQ6, the mean predicted coverage gain is 1.31 points against a realized 1.94 points. Removing the value weight from EPCG leaves 12 of the 15 combinations unchanged.

\subsubsection{Routing and solver statistics}
In CASH (all calls), the orchestrating agent selects Tool for 34.76\% and Agent for 65.24\% of account-days. In each evaluation period, the LP has one fractional account, and repair changes at most one account. Final allocations meet the predicted subsidy-rate constraint.

\subsubsection{Holm correction}
The Holm correction in RQ6 covers 36 paired MAE comparisons of Agent + EPCG against Random, EPCG, Rule + EPCG and Random50 + EPCG over three gaps and three feedback quotas, using 1,000 account-clustered paired bootstrap replicates.

\FloatBarrier
\section{Account-Level Effective Sample Size}
\label{app:kish}
Suppose the weights $\{W_i\}$ are fixed and the exceedance rates of accounts in a region are independent with common mean $\psi$ and variance $\psi(1-\psi)$, which corresponds to perfectly correlated observations within an account. The weighted mean $\hat\psi=\sum_iW_i\hat\psi_i/\sum_iW_i$ has variance
\begin{equation}
\mathrm{Var}(\hat\psi)=\psi(1-\psi)\frac{\sum_iW_i^2}{\big(\sum_iW_i\big)^2}=\frac{\psi(1-\psi)}{n^{\mathrm{eff}}_r},
\end{equation}
matching the variance of the mean of $n^{\mathrm{eff}}_r$ independent binary samples and motivating the pseudo-count in Eq.~\eqref{eq:beta}. With the same marginals and fixed weights, observations within an account that are not perfectly correlated give a smaller account-level variance, so the variance estimate with pseudo-count $n^{\mathrm{eff}}_r$ tends to be conservative. The derivation matches variances only and does not establish nominal coverage of the credible interval; bidding coupling across accounts, within-region heterogeneity and distribution drift can cause deviations. Since the true regional exceedance probability is not observable, we check on time slices the realized exceedance frequency of admitted regions in later cycles, the share of admissions followed by exceedance, and the agreement between predicted EPCG and realized coverage change.

\clearpage
\end{document}